\documentclass[iop,revtex4]{emulateapj}
\usepackage{natbib, color}
\usepackage[normalem]{ulem}
\usepackage{epstopdf}
\usepackage{graphicx}

\definecolor{red}{rgb}{1.0,0.0,0.0}

\newcommand{\Mj}[1]{$M_\mathrm{Jup}$}

\usepackage[utf8]{inputenc}

\begin{document}

\title{Dynamical Mass Measurement of the Young Spectroscopic Binary \\V343 Normae A\MakeLowercase{a}A\MakeLowercase{b} Resolved With the Gemini Planet Imager}
\author{Eric L. Nielsen\altaffilmark{1,2}, 
Robert J. De Rosa\altaffilmark{3}, 
Jason Wang\altaffilmark{3}, 
Julien Rameau\altaffilmark{4}, 
Inseok Song\altaffilmark{5},
James R. Graham\altaffilmark{3},
Bruce Macintosh\altaffilmark{2},
Mark Ammons\altaffilmark{6},
Vanessa P. Bailey\altaffilmark{2},
Travis S. Barman\altaffilmark{7},
Joanna Bulger\altaffilmark{8},
Jeffrey K. Chilcote\altaffilmark{9},
Tara Cotten\altaffilmark{5},
Rene Doyon\altaffilmark{4},
Gaspard Duch\^{e}ne\altaffilmark{3,10},
Michael P. Fitzgerald\altaffilmark{11},
Katherine B. Follette\altaffilmark{2},
Alexandra Z. Greenbaum\altaffilmark{12},
Pascale Hibon\altaffilmark{13},
Li-Wei Hung\altaffilmark{11},
Patrick Ingraham\altaffilmark{14},
Paul Kalas\altaffilmark{3},
Quinn M. Konopacky\altaffilmark{15},
James E. Larkin\altaffilmark{11},
J\'{e}r\^{o}me Maire\altaffilmark{9},
Franck Marchis\altaffilmark{1},
Mark S. Marley\altaffilmark{16},
Christian Marois\altaffilmark{17,18},
Stanimir Metchev\altaffilmark{19,20},
Maxwell A. Millar-Blanchaer\altaffilmark{21,9},
Rebecca Oppenheimer\altaffilmark{22},
David W. Palmer\altaffilmark{6},
Jenny Patience\altaffilmark{23},
Marshall D. Perrin\altaffilmark{24},
Lisa A. Poyneer\altaffilmark{6},
Laurent Pueyo\altaffilmark{24},
Abhijith Rajan\altaffilmark{23},
Fredrik T. Rantakyr\"o\altaffilmark{25},
Dmitry Savransky\altaffilmark{26},
Adam C. Schneider\altaffilmark{27},
Anand Sivaramakrishnan\altaffilmark{24},
Remi Soummer\altaffilmark{24},
Sandrine Thomas\altaffilmark{14},
J. Kent Wallace\altaffilmark{28},
Kimberly Ward-Duong\altaffilmark{23},
Sloane J. Wiktorowicz\altaffilmark{29},
Schuyler G. Wolff\altaffilmark{12}}

\altaffiltext{1}{SETI Institute, Carl Sagan Center, 189 Bernardo Avenue, Mountain View, CA 94043, USA}
\altaffiltext{2}{Kavli Institute for Particle Astrophysics and Cosmology, Stanford University, Stanford, CA 94305, USA}
\altaffiltext{3}{Astronomy Department, University of California, Berkeley, CA 94720, USA}
\altaffiltext{4}{Institut de Recherche sur les Exoplan\`{e}tes, D\'{e}partment de Physique, Universit\'{e} de Montr\'{e}al, Montr\'{e}al QC H3C 3J7, Canada}
\altaffiltext{5}{Department of Physics and Astronomy, University of Georgia, Athens, GA 30602, USA}
\altaffiltext{6}{Lawrence Livermore National Laboratory, 7000 East Ave., Livermore, CA 94550, USA}
\altaffiltext{7}{Lunar and Planetary Laboratory, University of Arizona, Tucson, AZ 85721, USA}
\altaffiltext{8}{Subaru Telescope, NAOJ, 650 North A’ohoku Place, Hilo, HI 96720, USA}
\altaffiltext{9}{Dunlap Institute for Astronomy \& Astrophysics, University of Toronto, 50 St. George St., Toronto, Ontario, Canada}
\altaffiltext{10}{University of Grenoble Alpes/CNRS, IPAG, F-38000 Grenoble, France}
\altaffiltext{11}{Department of Physics \& Astronomy, University of California, Los Angeles, CA 90095, USA}
\altaffiltext{12}{Department of Physics and Astronomy, Johns Hopkins University, Baltimore MD 21218, USA}
\altaffiltext{13}{European Southern Observatory , Alonso de Cordova 3107, Vitacura, Santiago, Chile}
\altaffiltext{14}{Large Synoptic Survey Telescope, 950 N Cherry Ave, Tucson AZ, 85719, USA}
\altaffiltext{15}{Center for Astrophysics and Space Sciences, University of California, San Diego, La Jolla, CA 92093, USA}
\altaffiltext{16}{Space Science Division, NASA Ames Research Center, Mail Stop 245-3, Moffett Field CA 94035, USA}
\altaffiltext{17}{National Research Council of Canada Herzberg, 5071 West Saanich Rd, Victoria, BC, V9E 2E7}
\altaffiltext{18}{University of Victoria, Department of Physics and Astronomy, 3800 Finnerty Rd, Victoria, BC V8P 5C2}
\altaffiltext{19}{Department of Physics and Astronomy, Centre for Planetary Science and Exploration, The University of Western Ontario, London, ON N6A 3K7, Canada}
\altaffiltext{20}{Department of Physics and Astronomy, Stony Brook University, Stony Brook, NY 11794-3800, USA}
\altaffiltext{21}{Department of Astronomy \& Astrophysics, University of Toronto, 50 St. George St., Toronto, Ontario, Canada}
\altaffiltext{22}{American Museum of Natural History, Depratment of Astrophysics, Central Park West at 79th Street, New York, NY 10024, USA}
\altaffiltext{23}{School of Earth and Space Exploration, Arizona State University, PO Box 871404, Tempe, AZ 85287, USA}
\altaffiltext{24}{Space Telescope Science Institute, 3700 San Martin Drive, Baltimore, MD 21218, USA}
\altaffiltext{25}{Gemini Observatory, Casilla 603, La Serena, Chile}
\altaffiltext{26}{Sibley School of Mechanical and Aerospace Engineering, Cornell University, Ithaca, NY 14853}
\altaffiltext{27}{Department of Physics and Astronomy, University of Toledo, 2801 W. Bancroft St., Toledo, OH 43606, USA}
\altaffiltext{28}{Jet Propulsion Laboratory, California Institute of Technology, 4800 Oak Grove Dr., Pasadena CA 91109, USA}
\altaffiltext{29}{The Aerospace Corporation, 2310 E. El Segundo Blvd., El Segundo, CA 90245}

\begin{abstract}
We present new spatially resolved astrometry and photometry from the Gemini Planet Imager of the inner binary of the young multiple star system V343 Normae, which is a member of the $\beta$ Pictoris moving group. V343 Normae comprises a K0 and mid-M star in a $\sim$4.5 year orbit (AaAb) and a wide $10\arcsec$ M5 companion (B). By combining these data with archival astrometry and radial velocities we fit the orbit and measure individual masses for both components of $M_{\rm Aa} = 1.10 \pm 0.10\ M_\odot$ and $M_{\rm Ab} = 0.290 \pm 0.018\ M_\odot$. Comparing to theoretical isochrones, we find good agreement for the measured masses and {\it JHK} band magnitudes of the two components consistent with the age of the $\beta$~Pic moving group.  We derive a model-dependent age for the  $\beta$~Pic moving group of $26 \pm 3$~Myr by combining our results for V343 Normae with literature measurements for GJ~3305, which is another group member with resolved binary components and dynamical masses. 
\end{abstract}


\maketitle

\section{Introduction}

Binaries remain the primary opportunity to directly measure the masses of stars. Resolved binaries provide the opportunity to simultaneously measure masses and individual fluxes of the components, and can be used to constrain atmospheric and evolutionary models (e.g.,  \citealt{muterspaugh:2008}, \citealt{schlieder:2016}, \citealt{dupuy:2014}, \citealt{crepp:2016}).  For binaries in young moving groups, where the age can be determined by considering all the stars in the group at once, the constraints placed on the models can be especially strong.

The $\beta$ Pic moving group was first identified by \citet{barrado1999}, and initially assigned an age of $20 \pm 10$~Myr, which was then revised to $12^{+8}_{-4}$~Myr by \citet{zucerkman:2001}.  More recent analyses of the group favor an older age:  $21 \pm 4$~Myr \citep{binks:2014},  $20 \pm 6$~Myr \citep{Macintosh:2015ew}, and $24 \pm 3$~Myr \citep{bell:2015}.  Stars in the moving group host the imaged planets $\beta$~Pic~b \citep{Lagrange:2012} and 51~Eri~b \citep{Macintosh:2015ew}, and brown dwarfs HR~7329~B \citep{lowrance:2000} and PZ~Tel~B \citep{biller:2010,mugrauer:2010}; the group is also home to the free-floating substellar object PSO J318.5338-22.8603 \citep{liu:2013}.  Since the inferred masses of these objects depend sensitively on their age, an accurate measurement of the age of the $\beta$ Pic moving group is of prime importance.

In addition to a number of wide binaries (e.g., \citealt{alonso:2015}), and close spectroscopic binaries such as HD 155555 AB \citep{bennett:1967} and V4046 Sgr \citep{byrne:1986} in the $\beta$ Pic moving group, there are two systems with resolved spectroscopic binaries---GJ~3305 and V343~Nor, which is resolved for the first time in this study. GJ 3305 is part of a triple system with the planet-host 51~Eri \citep{delorme:2012}, and an orbit with dynamical masses has recently been presented by \citet{montet:2015}, who find a model-dependent age for the system (and so the moving group) of $37 \pm 9$~Myr, consistent with previous estimates.  V343~Nor (HD~139084, HIP~76629, at a distance of $38.5^{+1.8}_{-1.6}$~pc \citep{vanLeeuwen:2007dc}), is a triple system consisting of a K0 primary and a mid-M secondary in a close binary orbit (V343 Nor Aa and Ab) and an outer $10\arcsec$ M5 companion (V343 Nor B; \citealp{song:2003}). An initial orbit fit to radial velocity (RV) measurements of V343 Nor A by \citet{thalmann:2014} found a 4.5-year orbit, eccentricity between 0.5 and 0.6, and a 0.11~$M_\odot$ minimum mass for the secondary.

The Gemini Planet Imager (GPI; \citealp{Macintosh:2014js}) is a near-infrared (NIR) integral field spectrograph and polarimeter at the Gemini South telescope.  As part of the ongoing GPI Exoplanet Survey (GPIES) we imaged V343~Nor A and resolved the Aa/Ab binary in 2015, and continued to monitor the system in 2016.  By combining these new astrometric epochs with archival imaging and RV data, we fit the orbit of the system and derive dynamical masses for both components. We derive a new estimate for the age of the $\beta$ Pic moving group based on stellar evolution models by combining our dynamical mass measurements with previous results for components of the GJ~3305 system. 

\section{Observations and Data Reduction}
\begin{figure*}
\centering
\includegraphics[trim={0 8.5cm 0 8cm},clip,width=1.0\textwidth]{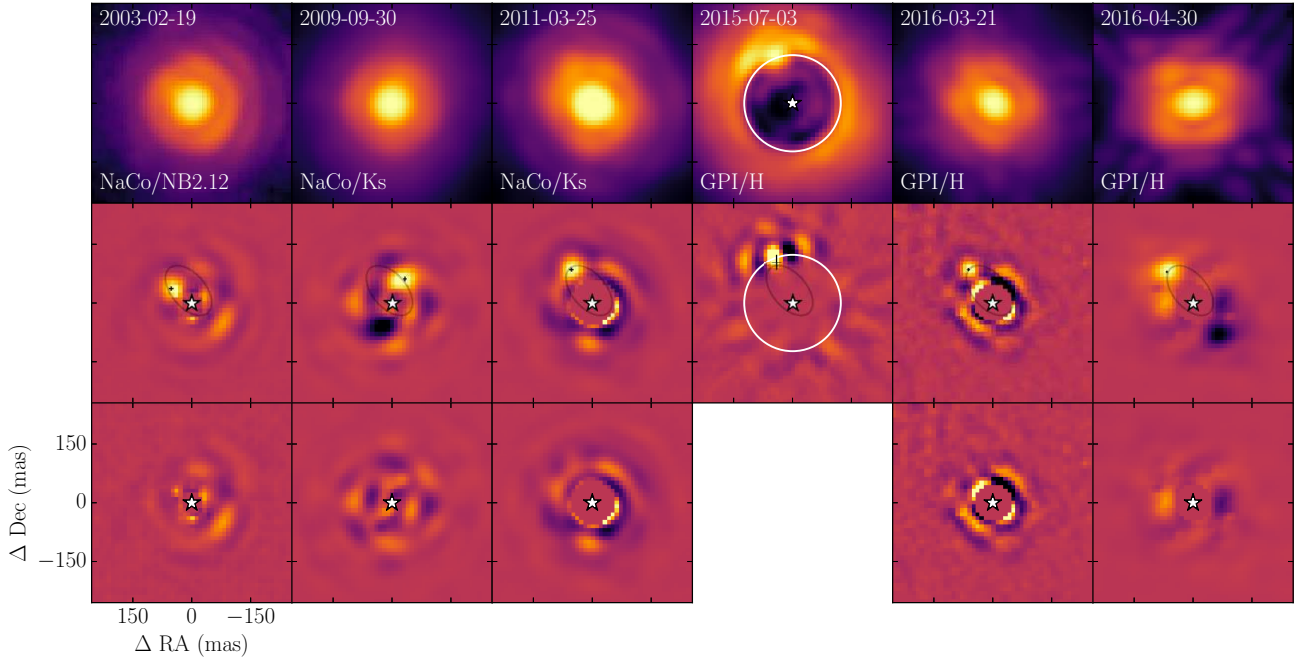}
\caption{V343 Nor Aa and Ab for the six epochs of imaging data.  The top row shows the initial reduced images, the middle row following PSF subtraction of the primary, and the bottom row the results of PSF fitting and subtraction of the secondary. Only two panels are shown for the July 2015 epoch as the secondary PSF was not subtracted. Labels at the top of each column indicate the epoch and the instrument. The best fit orbit (light grey) is indicated in the middle row, as is the derived location of the secondary ($+$). For the 2009 September and 2016 April epochs a negative of V343~Nor~Ab can be seen as the reference PSF was that of V343~Nor rotated by 180 degrees. These negatives do not appear in the final row as the companion was subtracted prior to the subtraction of the primary to create the final image. For the 2015 epoch, the white circle indicates the position and radius of the GPI coronographic mask.}
\label{fig:image}
\end{figure*}
\begin{deluxetable*}{cccccccccc}
\tabletypesize{\footnotesize}
\tablecaption{Astrometric and Photometric Measurements of V343 Nor Aa/Ab}
\tablewidth{0pt}
\tablehead{
\colhead{UT Date} &  \colhead{Instrument} & \colhead{Filter} & \colhead{Plate Scale} & \colhead{True North} & \colhead{$\rho$} & \colhead{$\theta$}  & \colhead{Contrast} & \colhead{Calib.}\\
&&&(mas px$^{-1}$)&(deg)&(mas)&(deg)&(mag)&Ref.}
\startdata
2003 Feb 19 & NaCo & NB2.12      & $13.24 \pm 0.05$   & $-0.05 \pm 0.10$ & 63.2 $\pm$ 6.8   & 54.6 $\pm$ 4.1  & -- & a\\
2009 Sep 30 & NaCo & $K_{\rm S}$ & $13.19 \pm 0.07$   & $-0.36 \pm 0.05$ & 68.0 $\pm$ 6.6   & 333.6 $\pm$ 3.0 & $2.86 \pm 0.11$ & b\\
2011 Mar 25 & NaCo & $K_{\rm S}$ & $13.21 \pm 0.02$   & $-0.50 \pm 0.05$ & 98.6 $\pm$ 4.6   & 32.6 $\pm$ 3.5  & -- & c\\
2015 Jul 03 & GPI  & $H$         & $14.166 \pm 0.007$ & $0.10 \pm 0.13$  & 111.3 $\pm$ 21.3 & 21.6 $\pm$ 5.0  & -- & d\\
2016 Mar 21 & GPI  & $H$         & $14.166 \pm 0.007$ & $0.10 \pm 0.13$  & 103.2 $\pm$ 4.3  & 35.8 $\pm$ 1.6  & -- & d\\
2016 Apr 30 & GPI  & $J$         & $14.166 \pm 0.007$ & $0.10 \pm 0.13$  & 100.5 $\pm$ 3.2  & 39.2 $\pm$ 1.3  & $3.25 \pm 0.12$ & d\\
2016 Apr 30 & GPI  & $H$         & $14.166 \pm 0.007$ & $0.10 \pm 0.13$  & 102.3 $\pm$ 3.3  & 40.4 $\pm$ 1.2  & $3.15 \pm 0.10$ & d
\enddata
\tablenotetext{a}{\citealp{Chauvin:2005dh}}
\tablenotetext{b}{\citealp{Vigan:2012jm}}
\tablenotetext{c}{\citealp{Chauvin:2015jy}}
\tablenotetext{d}{\citealp{DeRosa:2015jl}}
\label{tab:astrometry}
\end{deluxetable*}

\subsection{GPI Imaging}
V343~Nor~A was observed with GPI at three epochs: 2015 July 03, 2016 March 21, and 2016 April 30. The instrument configuration, observing conditions, and the fitting procedure used to measure the astrometry and photometry of the binary are described in more detail in the following subsections. The initial reduction steps for the data obtained at each epoch were broadly similar and are outlined first.

The raw images obtained at each epoch were processed with the GPI Data Reduction Pipeline (DRP) v1.3.0 \citep{Perrin:2014}, which performed dark current subtraction, bad pixel interpolation, and extraction of the microspectra within the raw 2-D image to create an $(x, y, \lambda)$ data cube \citep{Maire:2014gs}. The wavelength axis of the data cube was calibrated using observations of an argon arc lamp obtained immediately prior to each of the observing sequences. For the coronagraphic observations, the position of the star was determined by measuring the location of the satellite spots---attenuated replicas of the central point spread function (PSF) generated by a pupil-plane diffraction grating---within each wavelength slice of each reduced data cube \citep{Wang:2014}. For the observations taken in {\it unblocked} mode---where the focal plane mask is removed from the optical path---the position of the star was measured by fitting a Gaussian. Pixels with flux values exceeding the saturation limit during any read of the up-the-ramp calculation were flagged by the detector server, and excluded from the Gaussian fit. Each wavelength slice within each reduced cube was then registered to a common center using the measured position of the central star.

\subsubsection{2015 July}
\begin{figure}
\centering
\includegraphics[trim={0 2cm 0 2cm},clip,width=1.0\hsize]{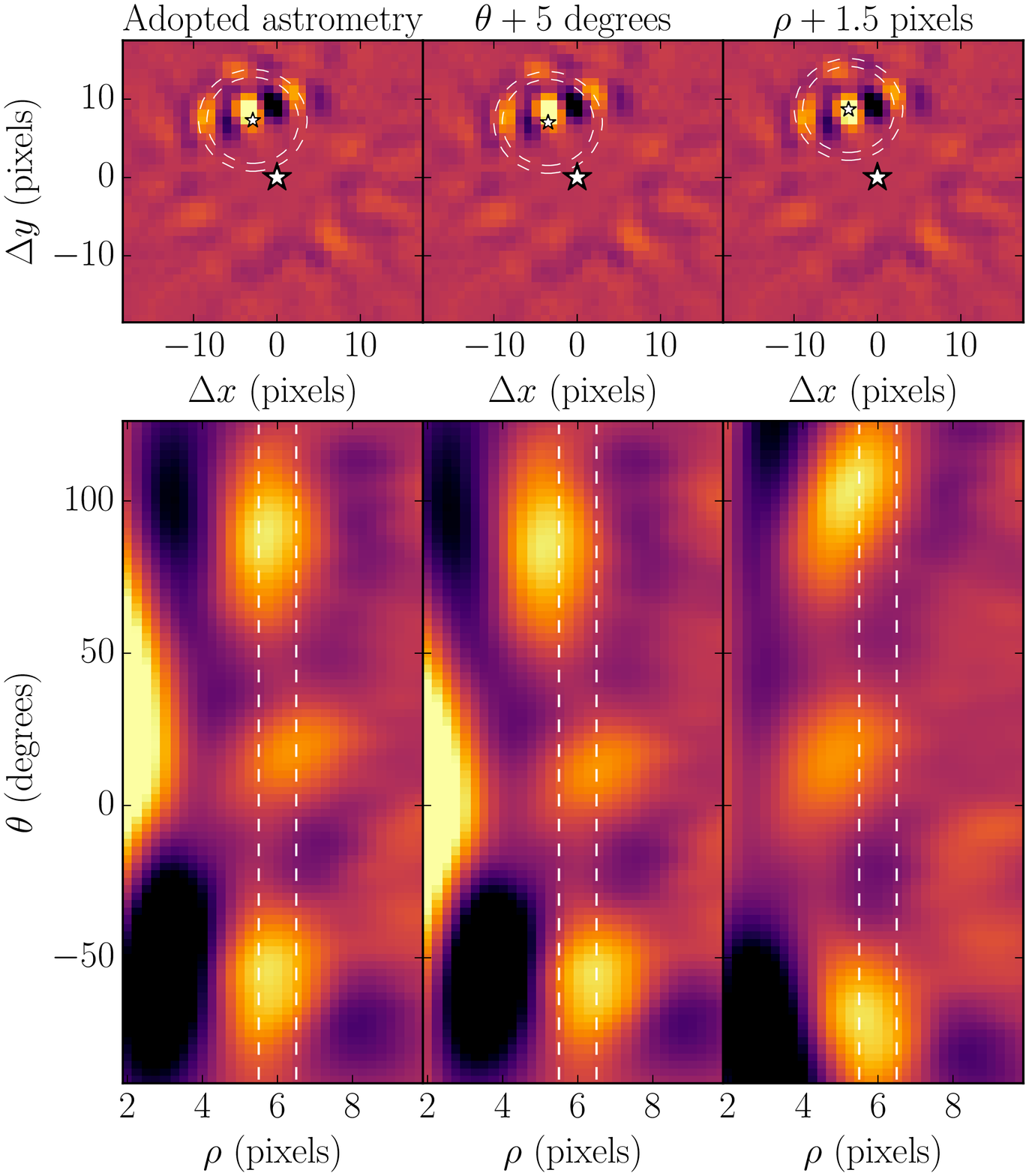}
\caption{PSF-subtracted wavelength-collapsed GPI images of V343~Nor~A at the 2015 July 03 epoch displayed in Cartesian coordinates (top row) and projected into polar coordinates centered on the tested location of the companion (bottom row). The three columns correspond to different tested locations of V343~Nor~Ab; the adopted astrometry given in Table~\ref{tab:astrometry} (left column), the same but with a five degree offset applied on the position angle (middle column), and instead with a 1.5~pixel offset applied on the separation (right column). In the top row, the position of the two components are indicated by the star symbols. The position of the companion was estimated by maximizing the flux within an annulus centered on the predicted location of the second Airy maximum (central radius of 6~px, and a width of 1~px; white dashed lines), which appears as a rectangular region within the projected images.}
\label{fig:airy_fitting}
\end{figure}

V343~Nor~A was observed with GPI as a part of the GPI Exoplanet Survey on 2015 July 03 (GS-2015A-Q-500). A total of 35 59.6-second single-coadd exposures were obtained at $H$ band in coronagraphic mode over the course of an hour, achieving a field rotation of $22\fdg2$. Observing conditions were worse than median, and deteriorated through the sequence, with an average seeing recorded by the Gemini DIMM telescope of $1\farcs14$. The data were reduced following the steps outlined previously. Half of the sequence was rejected due to poor image quality based on a visual inspection of the reduced data cubes. The remaining data cubes were further processed through an Angular Differential Imaging pipeline (ADI; \citealp{Marois:2006df}) in order to subtract the PSF of the primary star. After PSF subtraction, the individual wavelength slices from each data cube were rotated to align North with the vertical axis, and averaged together to create a final image.

Whilst the companion was apparent within the final image, it was partially occulted by the focal plane mask ($0\farcs246$ diameter at $H$) preventing a reliable measurement of its position using simple centroiding techniques. Instead, the position was measured by fitting the location of the second Airy maximum, located at a separation of $r\approx 6$~px from the center of the PSF. At each tested $(x, y)$ position for the companion, the image was projected into polar coordinates $(r, \theta)$, centered on the tested position of the companion. In the ideal case, when the tested location is coincident with the true location of the companion, the second Airy maximum will appear vertical within the projected image, being at a constant value of $r$ for all values of $\theta$ (Figure~\ref{fig:airy_fitting}, bottom left). As the tested location deviates from the true location of the companion, the Airy maximum no longer has a constant radius for all values of $\theta$, and becomes distorted within the projected image (Figure~\ref{fig:airy_fitting}, bottom middle and right). The location of the companion was thus estimated by maximizing the flux at the radii of the second Airy maximum ($r=6.5$~px) over all values of $\theta$ by varying the tested $(x, y)$ position. An uncertainty of $\sigma_{\rho} = 1.5$~px and $\sigma_{\theta} = 5$~deg was adopted. Deviations from the adopted astrometry greater than these uncertainties caused a significant shift in the position of the second Airy maximum within the projected image. The pixel offset between the two components was converted into an on-sky separation and position angle using the astrometric solution presented in \citet{DeRosa:2015jl}, and are given in Table~\ref{tab:astrometry}.

\subsubsection{2016 March \& April}
To obtain more accurate astrometry of the system, V343~Nor~A was observed with GPI in {\it unblocked} mode (GS-2015B-Q-501), where the focal plane mask which partially occulted the companion in the first epoch was removed from the optical path, at $H$ band on 2016 March 21, and at both $J$ and $H$ band on 2016 April 30. For the March epoch, seven 29.1-second single-coadd exposures were obtained under poor observing conditions and with significant extinction from clouds. For the April epoch, thirteen 1.5-second $\times$ 10-coadd frames were obtained, five at $J$ band and eight at $H$ band, under median observing conditions with negligible extinction.

PSF fitting was used to measure the pixel offset and flux ratio between the two components within each broadband image. As no unocculted PSF was observed either before or after the observations of V343~Nor~A, a reference PSF was constructed from either previous GPI observations of unocculted $H$-band PSFs from commissioning and the GPIES campaign, or from a 180~degree rotated version of the PSF of V343~Nor~A, based on the assumption that the PSF is rotationally symmetric (e.g., \citealp{Perrin:2003ff}). For each image from each epoch, the best matched reference PSF was that which, when fit to the position and flux of V343~Nor~Aa, minimized the sum of the residuals of the PSF-subtracted image within an annulus with a central radius of 7.3~px (the separation of V343~Nor~Ab) and width of 5~px. V343~Nor~Ab was masked within the annulus to minimize any bias to the fit. For the 2016 March epoch, the best fit reference PSF was identified as the PSF of the early M-dwarf companion to HIP~70931, observed during the commissioning of GPI on 2014 May 11. For the 2016 April epoch the PSF was elongated along one axis and as such the available reference PSFs were a poor match. Instead, a reference PSF was constructed for each broadband image by rotating the image by 180 degrees.

The pixel offset and flux ratio between the two components were measured within each broadband image by fitting simultaneously both components using the best fit reference PSF identified previously. For the March epoch six parameters were fit: the pixel position of the two components and the flux ratios between the reference PSF and each of the components. For the April epoch only five were fit, the flux ratio between the reference PSF and the primary star was set to unity as the reference PSF was a 180-degree rotated copy of the image. The parameters which minimized the residuals of the PSF-subtracted image within the same annulus described previously were found using a downhill simplex optimization algorithm. The results of the PSF subtraction are demonstrated for one of the images from each epoch in Figure~\ref{fig:image} (fifth and sixth columns, middle and bottom rows). Uncertainties were estimated by comparing the injected and recovered astrometry and photometry of fake companions injected at the same separation and flux ratio as V343~Nor~Ab, at seven different position angles with an interval of 45 degrees. This process was repeated for each of the images at a given epoch to determine a final pixel offset, flux ratio, and corresponding uncertainties for that epoch. As with the first epoch, these pixel offsets were converted into separations and position angles which are reported alongside the flux ratios in Table~\ref{tab:astrometry}.

\subsection{Archival NaCo Imaging}
To better sample the orbit of the V343~Nor~A binary, VLT/NaCo \citep{Rousset:2003hh,Lenzen:2003iu} observations were obtained from the ESO Archive\footnote{{\tt http://archive.eso.org/}}. The two components of the binary were resolved at three epochs between 2003 and 2011, of the six available within the archive. Observations were also obtained on the following dates but were not usable: 2003 Feb 14 (very poor conditions), 2012 Jul 21 (variable conditions, companion not resolved), and 2013 Feb 14 (taken with 27~mas~px$^{-1}$ camera, companion separation predicted to be $\sim$2~px). As with the GPI observations, the initial reduction steps were similar for each epoch, and are described below. A summary of the instrument configuration and observing mode, and of the fitting procedure used to measure the astrometry and photometry, is given for each epoch in the following subsections.

Each NaCo image was processed through the standard near-infrared data reduction process consisting of dark current subtraction, flat fielding, bad pixel interpolation, and sky subtraction. The location of the star within each image was measured by fitting a 2-D Gaussian, with non-linear and saturated pixels given zero weight. For observations taken in {\it cube} mode; where instead of multiple coadds being combined into one image, each coadd is saved to disk; which typically consist of thousands of raw images on a given target, the full width at half maximum (FWHM) of the central star was measured within each image, and the ten percent with the smallest FWHM were carried forward in this analysis. Reduced images were then registered to a common center, and high-pass filtered in the Fourier domain to remove spatial features larger than ten pixels. These reduced and aligned images were mean-combined to create a final image for each target.

\subsubsection{2003 February}
Twenty-eight images were obtained with NaCo on 2003 February 19 UT using the NB2.12 narrowband filter ($\lambda_{\rm cen} = 2.13$~\micron) and the S13 objective camera, each consisting of a 1.1-second $\times$ 2-coadd exposure (program ID 070.C-0777, PI: Mundt). Observing conditions were slightly worse than median, with an average DIMM seeing during the sequence of $0\farcs87$. Despite the narrowband filter and short exposure time, V343~Nor~Aa saturated the central core of the PSF to a radius of approximately two pixels, depending on the seeing conditions during the exposure. As described previously, such pixels were given zero weight when determining the position of the star.

The PSF within the observations of V343~Nor~A exhibited a strong asymmetry in the intensity of the first Airy ring, at radius coincident with that of V343~Nor~Ab, which precluded using a 180-degree rotated copy for PSF subtraction. Instead, we searched the ESO archive for a PSF reference star. Restricting the search to observations within a month of 2003 February 19, observations of seven stars were found and reduced. The best reference PSF was identified in a similar fashion as for the GPI observations, and was found to be that of GJ~256 (K4V; \citealp{Houk:1988wv}) observed as a part of the same program as V343~Nor~A. With the reference PSF identified, the PSF fitting and uncertainty estimation followed the same procedure outlined for the GPI observations. Pixel offsets were converted into an on-sky separation and position angle, reported in Table~\ref{tab:astrometry}, based on the astrometric calibration of NaCo S13 observations given in \citet{Chauvin:2005dh}. The saturation of the inner pixels within the core of the PSF prevented a measurement of the contrast between the two components at this epoch.

\subsubsection{2009 September}
Forty-one datasets were obtained in {\it cube} mode with NaCo on 2009 September 30 UT using the $K_{\rm S}$ broadband filter ($\lambda_{\rm cen} = 2.12$~\micron) and the S13 objective camera (program ID 083.C-0150, PI: Vogt), each with an integration time of 0.347 seconds, and between 120 and 122 coadds. In total, 4977 0.347-second exposures were recorded. Observing conditions were worse than median, with an average DIMM seeing of $1\farcs09$. During short periods of improved seeing, counts within the inner core of the PSF exceeded the linearity limit of the detector, although the saturation limit was never reached.

Despite the poor observing conditions the PSF within the combined image was symmetric, and as such a 180 degree rotated copy was used as a reference for PSF subtraction. Observations of other stars obtained with the $K_{\rm S}$ filter within a month were tested as reference PSFs, the residuals in each case were significantly worse. PSF fitting followed the same procedure as outlined previously. Pixel offsets were converted into an on-sky separation and position angle using the astrometric calibration given in \citet{Vigan:2012jm}, and are reported in Table~\ref{tab:astrometry}. As the pixel values exceeded the linearity limit in only a small subset of the images used within this analysis, a contrast between the two components was measured at this epoch, and is reported in Table~\ref{tab:astrometry}.

\subsubsection{2011 March}
Twenty-one datasets were obtained in {\it cube} mode with NaCo on 2011 March 25 UT using the $K_{\rm S}$ broadband filter ($\lambda_{\rm cen} = 2.12$~\micron) and the S13 objective camera (program ID 086.C-0600, PI: Vogt), each with an integration time of 0.347 seconds, and between 101 and 102 coadds. In total, 2135 0.347-second exposures were recorded. Observing conditions were better than median, with an average DIMM seeing of $0\farcs72$. Despite being observed with the same instrument configuration as the 2009 epoch, the improved seeing conditions caused the central core of the PSF to exceed the saturation limit of the detector within a three pixel radius. Saturated and non-linear pixels were given zero weight when determining the position of the central star within each image.

As with the previous two epochs, the best reference PSF was identified from either other stars observed with the $K_{\rm S}$ filter and S13 objective camera, or a 180 degree rotated copy of the PSF of V343~Nor~Aa. Fourteen potential reference PSF stars were identified within the archive. Of these fourteen, and the rotated copy of V343~Nor~Aa, the best reference PSF was identified as that of PZ~Tel (G9IV; \citealp{Torres:2006bw}), observed as a part of the same program as V343~Nor~A. The PSF fitting and uncertainty estimation followed the same procedure as previously.  The brown dwarf companion to PZ~Tel \citep{biller:2010} was at a separation of $0\farcs37$ at this epoch, wide enough to not bias the fit. Pixel offsets were converted into an on-sky separation and position angle using the astrometric calibration given in \citet{Chauvin:2015jy}, and are given in Table~\ref{tab:astrometry}. A measurement of the contrast between the two components could not be estimated from these observations due to the saturation of V343~Nor~Aa within a three pixel radius.

\subsection{Radial Velocities}

\subsubsection{HARPS} V343~Nor~A was observed by the $R\approx115000$ HARPS spectrograph \citep{mayor:2003} on the La Silla 3.6m telescope 14 times between 2005 and 2009.  We retrieved the pipeline-reduced spectra from the ESO Phase 3 spectral archive\footnote{{\tt http://archive.eso.org/wdb/wdb/adp/phase3\_spectral/form}}, which include barycentric radial velocities.  Typical errors on individual measurements are quoted as $<$5~m~s$^{-1}$, however by comparing points at similar epochs (within 1 week) we compute the standard deviation to determine that the stellar jitter is closer to 130~m~s$^{-1}$, and so we adopt this as the minimum RV error.  HARPS RVs, along with RVs from other sources, are reported in Table~\ref{tab:rv}.

\subsubsection{FEROS} There are 19 epochs of data on V343~Nor~A between 2002 and 2012 collected with the $R\approx48000$ FEROS spectrograph \citep{kaufer:1999} which was located on the ESO 1.52m telescope up to October of 2002, then moved to the MPG/ESO 2.2m telescope.  We have retrieved these data from the archive and reduced them with the standard MIDAS pipeline.  The data from 2004 and subsequent epochs include fibers from the star and a ThAr reference spectrum allowing for wavelength stabilization.  We measure relative radial velocities for these epochs by cross-correlating the 2005 March 23 epoch spectrum with the spectra from other epochs, producing relative RVs for all FEROS epochs since 2004.  Both FEROS and HARPS are wavelength-stabilized spectrographs with excellent RV stability over time, with internal errors much smaller than the published RVs of possible standards observed by FEROS on the same nights as the V343 Nor data.  So rather than convert the FEROS relative RVs to absolute RVs by comparing to standards, we instead convert these to absolute RVs on the same scale as the HARPS data by assigning an initial RV offset, and then fitting for this offset more precisely in the MCMC orbit fitting procedure (as described in Section~\ref{sec:orbit}).  The initial offset is chosen by assuming that the RV measured by FEROS on 2005 March 23 is identical to the RV measured from the 2005 April 01 HARPS epoch.  The values given in Table~1 use the more precise offset from the  $\chi^2_{\rm min}$ orbit fit.  We follow the same procedure as for the HARPS data, and assign an error of 160~m~s$^{-1}$ to these measurements, the standard deviation of RVs over FEROS epochs within two weeks of each other.

To extract an RV from the remaining FEROS epoch from 2002 April 04, before FEROS was moved to a different telescope, we cross correlate the V343~Nor~A spectrum with spectra from stars of similar spectral type and known RVs observed on the same night.  The standard deviation of errors from these standards is 80~m~s$^{-1}$, and so we assign the minimum error of 160~m~s$^{-1}$ to this point as well.

\subsubsection{UVES} V343~Nor~A was observed twice with UVES \citep{dekker:2000}, at VLT UT2, in 2007 and 2009, at $R\approx40000$.  We retrieve the reduced spectra from the ESO Phase 3 spectral archive and, as with the FEROS 2002 data, cross-correlate these spectra with spectra taken on the same night (or in the case of 2009, from nearby nights) of stars with measured RVs. Uncertainties are assigned as the scatter in the derived RV using all standards.

\subsubsection{SSO/LCO Echelle Spectra} As part of a large survey of nearby young stars \citep{song:2003}, V343~Nor~A was observed four times between 2001 and 2003 at Siding Spring Observatory (SSO) and Las Campanas Observatory (LCO). In 2001 and 2002, using an echelle spectrograph with the SSO 2.3m telescope, eight orders of the echelle spectra covering portions of the spectra between 5800 and 7250~\AA~were obtained. The measured spectral resolution of SSO spectra was $R\approx15000$ at orders containing the H$\alpha$ and Li\,$\lambda$6708 lines. In 2003, a $R\approx40000$ echelle spectrum was obtained with the LCO du Pont 2.5m telescope. These SSO/LCO spectra were cross-correlated against spectra of a handful of known RV standards and the typical RV uncertainties are 1.5~km~s$^{-1}$ and 0.5~km~s$^{-1}$ for SSO and LCO spectra, respectively.

\begin{deluxetable}{ccc}
\tabletypesize{\footnotesize}
\tablecaption{Radial velocities of V343 Nor Aa}
\tablewidth{0pt}
\tablehead{\colhead{UT Date} &  \colhead{Instrument} & \colhead{RV (km s$^{-1}$)}} 
\startdata
2001 Apr 06 & SSO & 2.2 $\pm$ 1.2 \\ 
2001 Jun 02 & SSO & 2.9 $\pm$ 2.2 \\ 
2002 Apr 04 & FEROS & 5.05 $\pm$ 0.16 \\
2002 Jul 25 & SSO & 5.4 $\pm$ 0.7 \\
2003 Mar 27 & LCO & 5.6 $\pm$ 0.4 \\ 
2004 Apr 04 & FEROS & -1.53 $\pm$ 0.16 \\
2005 Feb 19 & FEROS & 2.15 $\pm$ 0.16 \\
2005 Mar 23 & FEROS & 2.53 $\pm$ 0.16 \\
2005 Apr 01 & HARPS & 2.52 $\pm$ 0.13 \\
2005 Apr 05 & HARPS & 2.55 $\pm$ 0.13 \\
2005 Apr 07 & HARPS & 2.50 $\pm$ 0.13 \\
2005 May 09 & HARPS & 2.75 $\pm$ 0.13 \\
2005 May 31 & HARPS & 2.70 $\pm$ 0.13 \\
2006 Mar 11 & HARPS & 4.40 $\pm$ 0.13 \\
2006 Apr 02 & HARPS & 4.50 $\pm$ 0.13 \\
2006 Apr 03 & HARPS & 4.29 $\pm$ 0.13 \\
2006 May 05 & HARPS & 4.35 $\pm$ 0.13 \\
2006 May 18 & HARPS & 4.51 $\pm$ 0.13 \\
2006 May 22 & HARPS & 4.15 $\pm$ 0.13 \\
2006 May 24 & HARPS & 4.46 $\pm$ 0.13 \\
2006 Jun 10 & HARPS & 4.53 $\pm$ 0.13 \\
2009 Jul 20 & HARPS & 1.75 $\pm$ 0.13 \\
2007 May 02 & FEROS & 5.31 $\pm$ 0.16 \\
2007 May 06 & UVES & 5.5 $\pm$ 0.3 \\
2009 Mar 16 & UVES & 0.63 $\pm$ 1.0 \\
2012 Mar 05 & FEROS & 5.57 $\pm$ 0.16 \\
2012 Apr 14 & FEROS & 5.49 $\pm$ 0.16 \\
2012 Apr 15 & FEROS & 5.57 $\pm$ 0.16 \\
2012 Apr 16 & FEROS & 5.41 $\pm$ 0.16 \\
2012 Apr 16 & FEROS & 5.39 $\pm$ 0.16 \\
2012 Apr 18 & FEROS & 5.97 $\pm$ 0.16 \\
2012 Apr 22 & FEROS & 5.61 $\pm$ 0.16 \\
2012 Apr 23 & FEROS & 5.30 $\pm$ 0.16 \\
2012 Apr 26 & FEROS & 5.33 $\pm$ 0.16 \\
2012 Jul 03 & FEROS & 5.12 $\pm$ 0.16 \\
2012 Jul 03 & FEROS & 4.97 $\pm$ 0.16 \\
2012 Jul 10 & FEROS & 4.80 $\pm$ 0.16 \\
2012 Jul 10 & FEROS & 4.66 $\pm$ 0.16 \\
2012 Jul 13 & FEROS & 4.92 $\pm$ 0.16 \\
2012 Jul 13 & FEROS & 4.91 $\pm$ 0.16
\enddata
\label{tab:rv}
\end{deluxetable}

\begin{deluxetable}{lcc}
\tabletypesize{\footnotesize}
\tablecaption{Orbital Parameters of V343 Nor Aa/Ab}
\tablewidth{0pt}
\tablehead{\colhead{Parameter} &  \colhead{$\chi^2_{\rm min}$} & \colhead{Posterior}}
Semimajor axis ($a$, au) & 3.076 & 3.07 $\pm$ 0.08 \\
Eccentricity ($e$) & 0.5306 & 0.534 $\pm$ 0.019 \\
Inclination Angle ($i$, deg) & 54.87 & 55 $\pm$ 4 \\
Argument of Periastron ($\omega$, deg) & 129.18 & 130 $\pm$ 3 \\
Position Angle of Nodes ($\Omega$, deg) & 49.25 & 49.4 $\pm$ 1.6 \\
Epoch of Periastron Passage ($T_0$) & 2008.404 & 2008.41 $\pm$ 0.04 \\
Period ($P$, yr) & 4.576 & 4.58 $\pm$ 0.02 \\
Total Mass ($M_{\rm tot}$, $M_{\odot}$) & 1.391 & 1.39 $\pm$ 0.11 \\
Primary Mass ($M_1$, $M_{\odot}$) & 1.177 & 1.10 $\pm$ 0.10 \\
Secondary Mass ($M_2$, $M_{\odot}$) & 0.2875 & 0.290 $\pm$ 0.018 \\
System RV (RV$_0$, km~s$^{-1}$) & 2.920 & 2.92 $\pm$ 0.08 \\
FEROS Offset (RV$_{\rm F}$, km~s$^{-1}$) & 0.1782 & 0.20 $\pm$ 0.11 \\
\enddata
\label{tab:orbit}
\end{deluxetable}

\section{Analysis}

\subsection{Orbit Fitting}\label{sec:orbit}

\begin{figure*}
\centering
\includegraphics[trim={0 8cm 0 0},clip,width=1.0\textwidth]{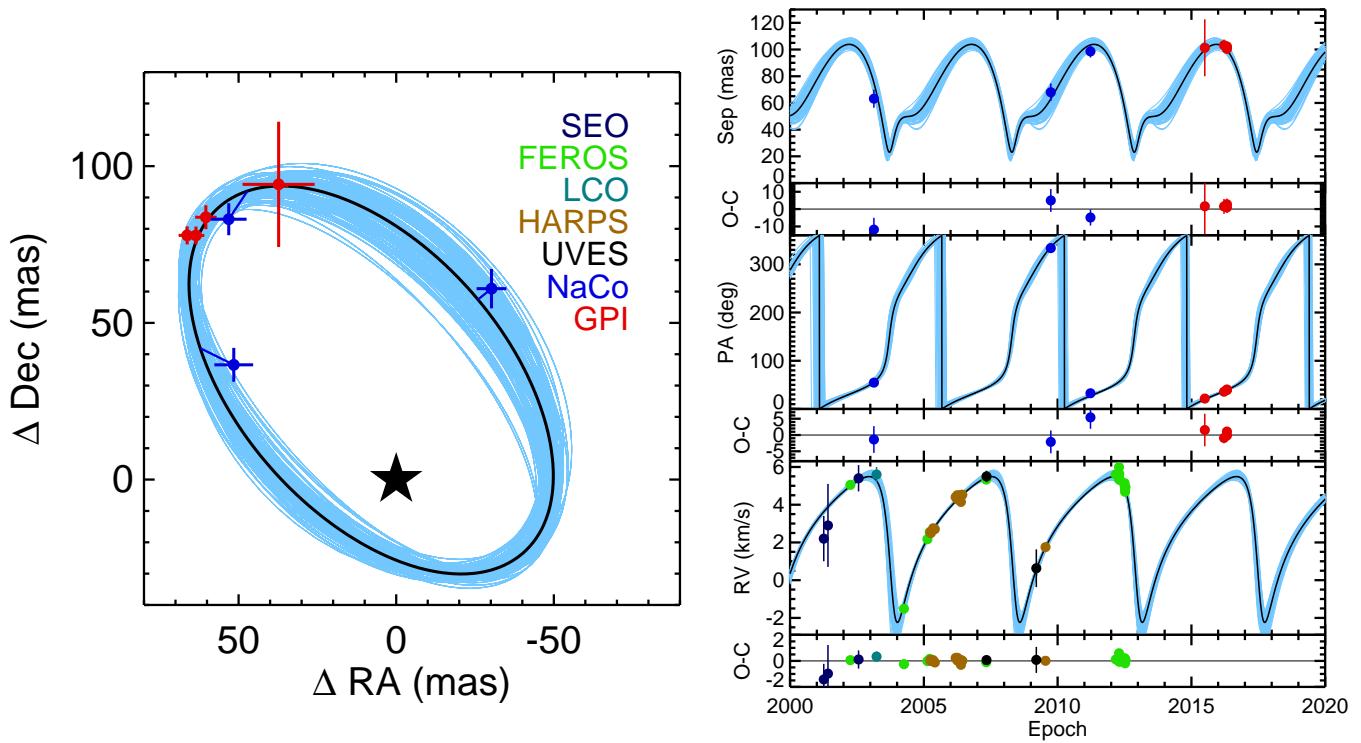}
\caption{
The best-fitting orbit (black) and 100 randomly-selected orbits (blue) from the posterior distribution.  The visual orbit is shown on the left, with solid lines connecting each data point to the location on the best-fitting orbit corresponding to the observing epoch.  On the right are separation, position angle, and RV against time, with Observed-Calculated (O-C) residuals given below each plot, with respect to the best-fitting orbit.  The combination of astrometry and RVs together provide a well-constrained orbit.
}
\label{fig:orbit}
\end{figure*}

\begin{figure*}
\centering
\includegraphics[width=1.0\textwidth]{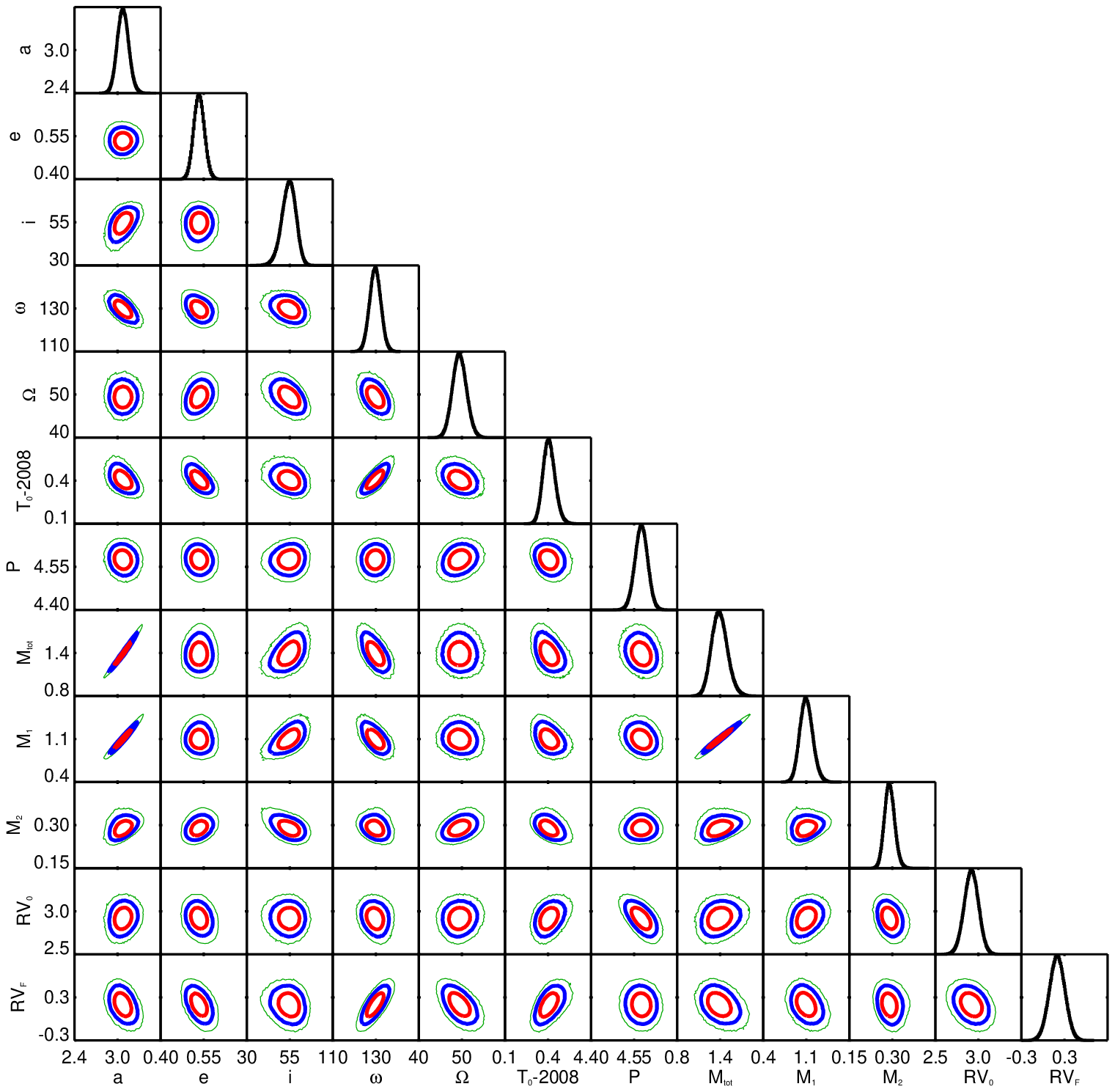}
\caption{Marginalized posterior distributions on all orbital parameters in diagonal elements, and covariances between them in the off-diagonal elements, with red, blue, and green contours representing 68\%, 95\%, and 99.7\% enclosed probability.  Fit orbital parameters are semimajor axis ($a$, au), eccentricity ($e$), inclination angle ($i$, degrees), argument of periastron ($\omega$, degrees), position angle of nodes ($\Omega$, degrees), epoch of periastron passage (T$_0$, decimal years), total mass ($M_{\rm tot}$, solar masses), secondary mass ($M_{2}$, solar masses), system RV (RV$_0$, km~s$^{-1}$), and FEROS RV offset (RV$_{\rm F}$, km~s$^{-1}$).  Also included are the derived quantities of period (P, years) and primary mass ($M_1$, solar masses). Covariances indicate that future astrometric observations that better constrain the semimajor axis will improve the errors on the mass of V343 Nor Aa.}
\label{fig:triangle}
\end{figure*}

We fit the combined astrometry and RVs using a Metropolis-Hastings Markov Chain Monte Carlo (MCMC) procedure previously described in \citet{Nielsen:2014}. Ten chains are run in parallel for $10^7$ steps each, running a 10-element fit for semimajor axis ($a$), eccentricity ($e$), inclination angle ($i$), argument of periastron ($\omega$), position angle of nodes ($\Omega$), epoch of periastron passage ($T_0$), total mass ($M_{\rm tot}$), secondary mass ($M_2$), system RV (RV$_0$), and FEROS RV offset (RV$_{\rm F}$).  We note that the final term, the offset between the FEROS and HARPS RVs, is not meant to be a measurement of the offset between the zero-points of the two instruments, but rather is a combination of the zero-point offset, RV jitter between the 2005 April 01 HARPS point and the 2005 March 23 FEROS point, individual measurement errors, and orbital motion between those two epochs.  Period ($P$) is then derived from semimajor axis and total mass using Kepler's third law and the distance to the system, and primary mass ($M_1$) from the difference between total mass and secondary mass.  We have found this choice of fit and derived parameters leads to faster convergence, and has a more natural set of priors. Priors are flat in $\log(a)$ and $\cos(i)$, and flat in all other parameters.  The chains are fully converged, producing Gelman-Rubin statistics less than 1.002 for all parameters.  Figure~\ref{fig:orbit} displays the data, the lowest $\chi^2$ orbit, and 100 representative orbits drawn from the posterior parameter distributions.  In Figure~\ref{fig:triangle} we present the posteriors on the orbital parameters and covariances between them.

We confirm the orbital period and eccentricity reported by \citet{thalmann:2014} based only on RV data, a significant subset of which appears within this analysis. The orbit is somewhat inclined ($i = 55 \pm 4^\circ$) and eccentric ($e = 0.534 \pm 0.019$), with a well-defined period of $P = 4.58\pm0.02$ yrs ($1671 \pm 8$ days).  The relative dearth of astrometric data means the fractional error on semimajor axis is significantly larger than the error on the period, so the total mass, and thus the primary mass, is only constrained at the 9\% level: $1.10 \pm 0.10$~$M_\odot$.  The precise RV observations, however, allow us to constrain the secondary mass to 6\%: $0.290 \pm 0.018$~$M_\odot$.

The system RV of $2.92 \pm 0.08$ km~s$^{-1}$ is a significant improvement over the value of $0.5 \pm 0.9$ km~s$^{-1}$ used in \citet{zucerkman:2001}.  Despite being over 2$\sigma$ from the initial value, the updated {\it UVW} velocity of the V343 Nor system still places it solidly in the $\beta$ Pic moving group.  We find a new {\it UVW} of the V343~Nor system of $U=-9.38 \pm 0.53$, $V=-17.27 \pm 0.70$, and $W=-9.84 \pm 0.48$~km~s$^{-1}$, within 1-$\sigma$ of the \citet{gagne2014} $\beta$~Pic moving group values of $[-11.03 \pm 1.38, -15.61 \pm 1.56, -9.24 \pm 2.50]$~km~s$^{-1}$, as expected for a group member.

\subsection{Comparison to Models}

\begin{figure*}
\centering
\includegraphics[trim={0 5.5cm 0 0},clip,width=1.0\textwidth]{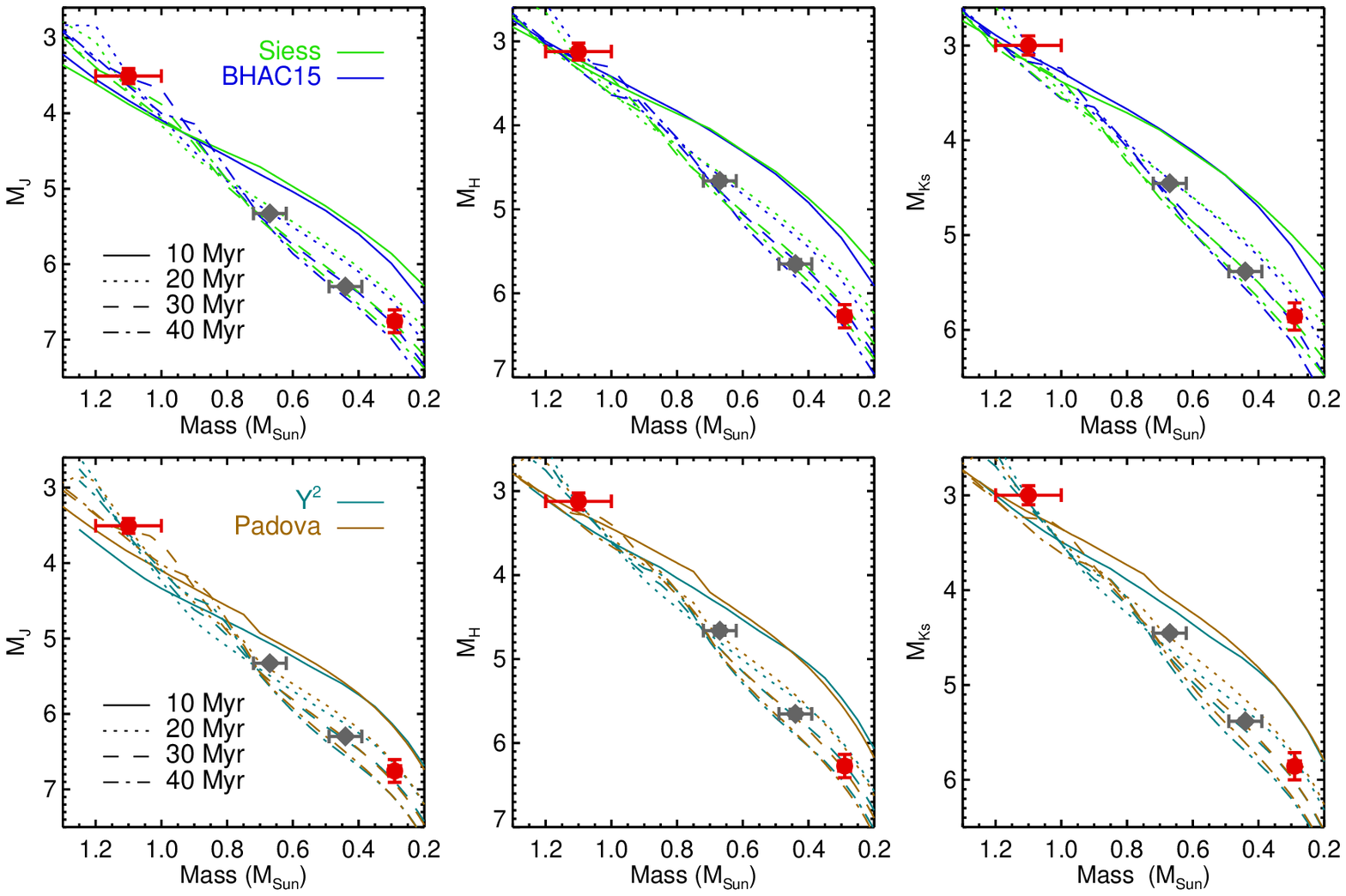}
\caption{Comparison of the measured photometry and masses of V343 Nor Aa/Ab (red circles) to the four sets of theoretical models, as well as GJ 3305 A/B (gray diamonds) from \citet{montet:2015}, showing good agreement for the expected 20--30 Myr age of the system.  For the plotted isochrones, colors corresponds to different models, and linestyles to ages.}
\label{fig:model}
\end{figure*}

With direct measurements of the mass of both V343~Nor~Aa and Ab, {\it JHK} photometry, and an age for the system of $24 \pm 3$~Myr \citep{bell:2015}, we can test the accuracy of theoretical stellar models with our observations of the system.  We begin with the 2MASS combined photometry of the system \citep{skrutskie:2006}, the {\sl Hipparcos} distance to V343 Nor Aa \citep{vanLeeuwen:2007dc}, and our measured contrasts between the two stars.  Utilizing the associated error with each quantity we use Monte Carlo error analysis to derive errors on apparent and absolute resolved magnitudes while accounting for the correlated errors.  In Figure~\ref{fig:model} we plot the absolute {\it JHK} magnitudes and dynamical masses against isochrones of the Siess \citep{siess:2000}, BHAC15 \citep{baraffe:2015}, Yonsei-Yale (Y$^2$, \citealt{spada:2013}), and Padova PARSEC \citep{bressan:2012} model grids.  Detailed discussion of the differences and similarities between these model grids, including comparison to a larger sample of benchmark stars, is presented in \citet{stassun2014} and \citet{herczeg2015}.  In addition, we include photometry and dynamical masses of GJ 3305  \citep{montet:2015} the only other resolved spectroscopic binary within the $\beta$~Pic moving group.  For all four objects the 1$\sigma$ errors lie between the 20 and 30 Myr isochrones, showing good agreement between the theoretical predictions and the measurements in the NIR for the expected age of the $\beta$ Pic moving group, over almost a factor of 4 in mass.

\subsection{The Age of the $\beta$ Pic Moving Group}\label{agesec}

\begin{figure}
\centering
\includegraphics[trim={0 8.0cm 0 0},clip,width=1.0\hsize]{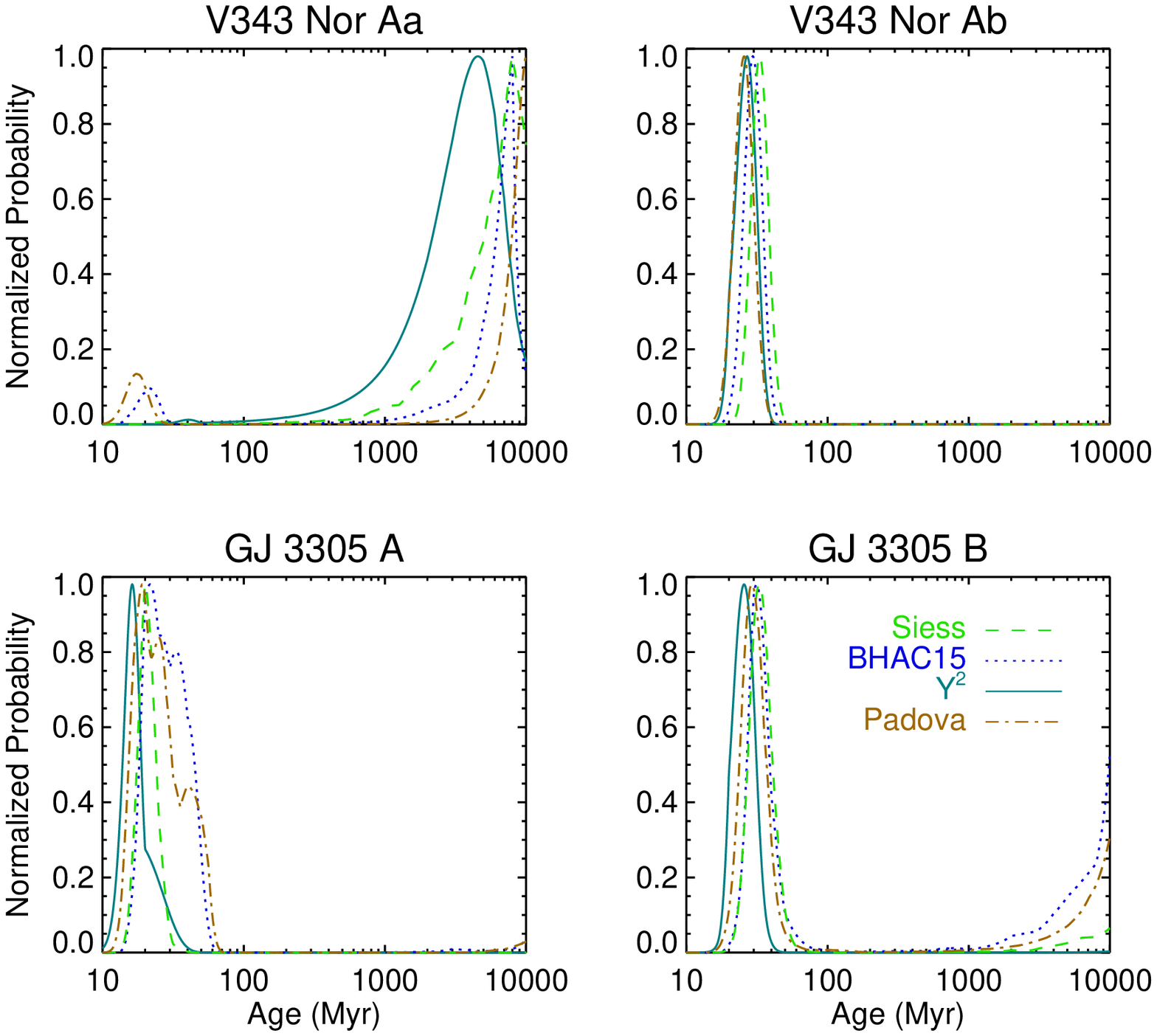}
\caption{
Normalized posterior distributions for age based on isochrones, measured mass, 
and photometry of V343 Nor Aa/Ab and GJ 3305 A/B.  When the measurement touches the main sequence, the uniform age prior favors large ages, but a young age is the clear best fit for V343 Nor Ab, GJ 3305 A, and GJ 3305 B, despite the tail at older ages.}
\label{fig:model2}
\end{figure}

We proceed to derive model-dependent ages for the $\beta$~Pic moving group based on these four members with dynamical masses.  We use a similar Bayesian method as in \citet{nielsen:2013} to construct posterior probability distributions for age.  We construct a three-dimensional grid uniformly sampled in distance and mass and logarithmically sampled in age.  Each point in the grid contains {\it JHK} apparent magnitudes by linearly interpolating the isochrones.  We then compute $\chi^2$ given the measured photometry and errors in each band, and likelihood is taken as $Prob. \propto e^{-\chi^2 /2}$, as we assume Gaussian errors.  Priors are flat in linear age (given a uniform star formation rate), and Gaussians in parallax and mass, corresponding to the measurements and 1$\sigma$ uncertainties.  Final age posterior probability density functions (PDFs) are computed by marginalizing over mass and distance, and are plotted in Figure~\ref{fig:model2}

For the M stars, GJ 3305 A and B and V343 Nor Ab, the NIR fluxes from the isochrones are monotonically decreasing with age over the initial tens of Myr, leading to well-defined peaks in probability between 10 and 40 Myr (Figure~\ref{fig:model2}).  A prior that is uniform in age assigns a probability of an age between 1 and 10 Gyr that is 100 times larger than for an age between 1 and 100 Myr.  As a result, even though the photometry of GJ~3305~A and B is most consistent with a $\sim$20~Myr age, there is a slight rise in probability for these two stars at several Gyr since the high-mass end of the 1-$\sigma$ confidence intervals reaches close to the main sequence.  For the solar-mass star V343 Nor Aa the 1-$\sigma$ errors in mass are consistent with the zero age main sequence (ZAMS), so an age of several Gyr is preferred as this is where solar-type stars stay for the majority of their lives.  However, the age posteriors show an increase in probability at $\sim$20 Myr reflecting the temporary increase in luminosity during the pre-main sequence evolution expected as the core becomes radiative and core nuclear burning begins, but before the core expands and the luminosity decreases toward the ZAMS.

\begin{figure}
\centering
\includegraphics[width=1.0\hsize]{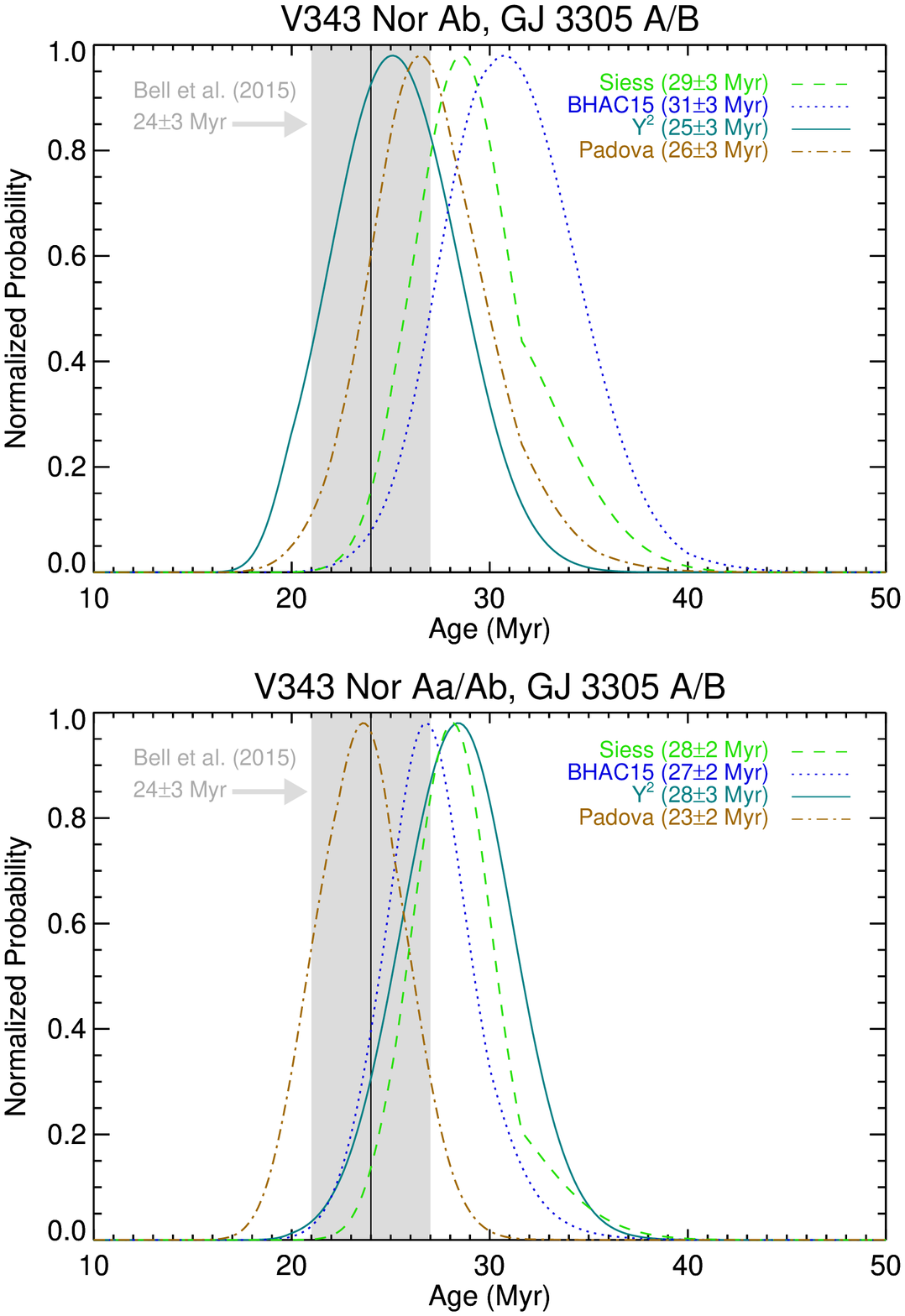}
\caption{
Model dependent age posterior probability distributions for the $\beta$ Pic moving group derived from the three M stars (top) and from all four stars including V343 Nor Aa (bottom).  The PDFs show well-constrained ages, and are consistent with the $24 \pm 3$~Myr age for the moving group as a whole derived by \citet{bell:2015}, indicated by a vertical line and gray region for the 1-$\sigma$ uncertainty.  By combining all the PDFs together we reach a final age for the $\beta$ Pic moving group of $26 \pm 3$~Myr.}
\label{fig:model3}
\end{figure}

In order to produce model-dependent ages of the $\beta$ Pic moving group we treat the four stars as independent measurements of the age, and multiply the posterior PDFs together.  The PDFs are not truly independent given the covariances between the binaries in mass, NIR fluxes, and distance, however given the similarity in the location of the probability peaks this should be a minor effect.  The top panel of Figure~\ref{fig:model3} contains the product of the PDFs for only the M stars, showing generally similar PDFs for each of the four models, with ages ranging from $25 \pm 3$~Myr for the Y$^2$ models to $31 \pm 3$~Myr for the BHAC15 models.   Ages are generally older than the $24 \pm 3$ Myr derived by \citet{bell:2015}, but overall consistent.  The bottom panel includes V343 Nor Aa as well, where the models' treatment of the onset of a radiative, nuclear-burning core and the subsequent rise in luminosity adds additional constraints on the age.  In general the widths of the posteriors shrinks upon using all four stars, and most models move toward a younger age, with the different models producing ages from $23 \pm 2$~Myr for the Padova models to $28 \pm 3$~Myr for the Y$^2$ models.  The Padova models show good agreement with the age determination of \citet{bell:2015}, while the other three models are at higher ages, but consistent at the $\sim$1-$\sigma$ level.

To compute a final age for the $\beta$ Pic moving group based on both sets of binaries with astrometric masses we average the four model-dependent posterior PDFs.  This gives each set of isochrones equal weight, and partially accounts for both the measurement error and the systematic error of the different models.  Our final age determination of the $\beta$ Pic moving group based on these dynamical masses is then 26$ \pm $3 Myr.

\section{Discussion}

Our age of $26 \pm 3$~Myr is slightly older than the $24 \pm 3$~Myr age of \citet{bell:2015}, which is derived from isochrone fitting to all the stars in the group, but within 1-$\sigma$ uncertainties the two ages are consistent with each other.  While the differences between the models continues to be the primary source of uncertainty in these ages, more precise measurements of the photometry of the components of V343 Nor A, as well as more constrained masses for all objects, will improve the precision of the age determinations for each individual model.  Much of the uncertainty in the mass of V343 Nor Aa comes from the uncertainty in the orbital semimajor axis (as seen in the strong covariance between the two in Figure~\ref{fig:orbit}), so future astrometric monitoring of the orbit over the next 1--2 years will be necessary to reduce these mass uncertainties.

Ages for the $\beta$ Pic moving group have been derived through a number of different methods, each converging on similar values in recent years.  \citet{binks:2014}, \citet{malo:2014}, and \citet{Macintosh:2015ew} utilized the lithium depletion boundary to derive ages of $21 \pm 4$ Myr,  $26 \pm 3$~Myr, and $20 \pm 6$~Myr respectively.  \citet{mamajek:2014} find a consistent traceback age for the system, but with large uncertainties, and an isochronal age of $23 \pm 3$~Myr.  The most recent contribution by \citet{bell:2015} updates the isochronal age to $24 \pm 3$~Myr.  \citet{montet:2015} found an age for GJ 3305 of $37 \pm 9$~Myr from a Bayesian analysis of the dynamical masses and resolved and integrated photometry of the system (we find similar results when using the BCAH15 models and only GJ 3305 AB $JHK$ photometry, deriving an age of 32$ \pm $6~Myr).  Our age of $26 \pm 3$~Myr, based on isochrone fitting only to stellar members with dynamical masses, is well within the ranges of these measurements, and gives further weight to a $\sim$25~Myr age of the $\beta$~Pic moving group.  \citet{malo:2014} use the spatial extent of the $\beta$ Pic moving group to estimate a time scale of star formation across the group of $\sim$5 Myr, similar in scale to the uncertainty we and other authors present.  A direct measurement of the age dispersion in the group will likely require a higher degree of age precision for individual stars than can currently be reached.

The masses of the directly imaged planets 51 Eri b and $\beta$ Pic b are largely unchanged if we adopt this new age.  Using the BT-SETTL models with the \citet{Caffau:2011ik} solar abundances\footnote{{\tt https://phoenix.ens-lyon.fr/Grids/BT-Settl/CIFIST2011}} (\citealt{baraffe:2015}, Allard et al. 2016 in prep) we compute masses by linearly interpolating the grid for absolute $H$ magnitude only.  Errors are computed using a Monte Carlo procedure taking into account errors in age and absolute $H$.  When moving from $20 \pm 6$~Myr of \citet{Macintosh:2015ew} to the $26 \pm 3$~Myr used here (and using the \citealt{Macintosh:2015ew} value for $H$ magnitude) the mass of 51~Eri~b increases from $2.3 \pm 0.5$~$M_{\rm Jup}$ to $2.7 \pm 0.3$~$M_{\rm Jup}$.  For $\beta$~Pic~b, when using the $H$ magnitude and age of $21 \pm 4$~Myr from \cite{morzinski:2015}, the mass increases from $12.4^{+0.9}_{-0.8}$~$M_{\rm Jup}$ to $13.4^{+2.9}_{-1.0}$~$M_{\rm Jup}$.

\section{Conclusion}

We have presented new imaging observations of V343 Nor A, as well as archival imaging of the pair and RVs of the primary, which taken together allow for a precise determination of the orbit.  Our orbit fit shows well-defined orbital parameters, including dynamical masses of $1.10 \pm 0.10$~$M_\odot$ and $0.292 \pm 0.018$~$M_\odot$ for the two components.  The V343 Nor A system thus joins GJ 3305 as only the second resolved spectroscopic binary with dynamical masses in the $\beta$ Pic moving group that can serve as a benchmark for testing models of pre-main sequence evolution.

Future astrometric monitoring of this system will further improve the mass precision and thus the precision of the model-dependent age of the $\beta$ Pic moving group.  At present there is no astrometric measurement of the system when the projected separation drops below 60 mas. Observations in 2017, when the projected separation should be between 20 and 40 mas, will be especially helpful in improving the precision of the mass measurements.  Such close separations should be reachable with Non Redundant Masking (NRM) observations, and is an ideal target for GPI NRM \citep{greenbaum:2014}.  2017 will also see the rapid decrease in RV from 6 to -2~km~s$^{-1}$, a poorly sampled part of the RV phase curve. This system also makes an excellent target for NIR spectroscopic observations to detect the spectral lines of the secondary, making the system a double-lined spectroscopic binary. Such measurements would further constrain the orbital parameters, specifically the mass ratio of the system (e.g., \citealp{Mazeh:2003eo}).

Overall there is excellent agreement between the age of the $\beta$ Pic moving group, as derived from isochrone fitting to all the stars \citep{bell:2015}, the theoretical models, and the dynamical masses and NIR photometry of these objects spanning a factor of $\sim$4 in stellar mass.  Identification and monitoring of new resolved $\beta$ Pic moving group binaries with short enough orbital periods to provide dynamical masses on reasonable timescales can further test the reliability of the models' relations between mass, age, and photometry.

\acknowledgments
We thank the anonymous referee for the helpful comments that improved the quality of this work.  Based on observations obtained at the Gemini Observatory, which is operated by the Association of Universities for Research in Astronomy, Inc., under a cooperative agreement with the National Science Foundation (NSF) on behalf of the Gemini partnership: the NSF (United States), the National Research Council (Canada), CONICYT (Chile), the Australian Research Council (Australia), Minist\'{e}rio da Ci\^{e}ncia, Tecnologia e Inova\c{c}\~{a}o (Brazil) and Ministerio de Ciencia, Tecnolog\'{i}a e Innovaci\'{o}n Productiva (Argentina). This research has made use of the SIMBAD database, operated at CDS, Strasbourg, France. Supported by NSF grants AST-0909188 and AST-1313718 (R.J.D.R., J.R.G., J.J.W., T.M.E., P.G.K.), AST-1411868 (B.M., K.F., J.L.P., A.R., K.W.D.), AST-141378 (P.A., G.D., M.P.F.), NNX11AF74G (A.Z.G., A.S.), and DGE-1232825 (A.Z.G.). Supported by NASA grants NNX15AD95G/NEXSS and NNX11AD21G (R.J.D.R., J.R.G., J.J.W., T.M.E., P.G.K.), and NNX14AJ80G (E.L.N., S.C.B., B.M., F.M., M.P.). J.R., R.D. and D.L. acknowledge support from the Fonds de Recherche du Qu\'{e}bec. Portions of this work were performed under the auspices of the U.S. Department of Energy by Lawrence Livermore National Laboratory under Contract DE-AC52-07NA27344 (S.M.A.). B.G. and M.J.G. acknowledge support from the National Sciences and Engineering Research Council of Canada. G.V. acknowledges a JPL Research and Technology Grant for improvements to the GPI CAL system.

{\it Facility:} \facility{Gemini:South (GPI)}.

\end{document}